\documentclass[prb,twocolumn,floatfix,groupeaddress]{revtex4}
\usepackage{amsmath,amsfonts,amssymb}
\usepackage{graphicx}
\usepackage{color}
\usepackage{enumerate}
\usepackage{verbatim}

\begin{document}

% -------------------------------------------------------------------

\title{Non-Equilibrium Charge Dynamics in Majorana-Josephson Devices}

\author{Ian J. van Beek,$^1$ Alfredo Levy Yeyati,$^2$ and Bernd Braunecker$^1$}

\affiliation{\textsuperscript{1} \hspace{-0.5em} SUPA, School of Physics and Astronomy,
	University of St Andrews, North Haugh, St Andrews KY16 9SS, UK \\
	\textsuperscript{2} \hspace{-0.5em} Departamento  de  F\'{i}sica  Te\'{o}rica  de  la  Materia  Condensada, Condensed  Matter  Physics Center (IFIMAC)  and  Instituto  Nicol\'{a}s  Cabrera,  Universidad  Aut\'{o}noma  de  Madrid, 28049 Madrid,  Spain}

\date{\today}

% -------------------------------------------------------------------

\begin{abstract}
We investigate the impact of introducing Majorana bound states, formed by a proximitized semiconducting nanowire in the topological regime, into a current biased capacitive Josephson junction, thereby adding delocalized states below the superconducting gap. We find that this qualitatively changes the charge dynamics of the system, diminishing the role of Bloch oscillations and causing single-particle tunnelling effects to dominate. We fully characterize the resulting charge dynamics and the associated voltage and current signals. Our work reveals a rich landscape of behaviours in both the static and time-varying driving modes. This can be directly attributed to the presence of Majorana bound states, which serve as a pathway for charge transport and enable non-equilibrium excitations of the Majorana-Josephson device.
\end{abstract}

% -------------------------------------------------------------------

\maketitle

% -------------------------------------------------------------------
\section{Introduction}

In recent years, there has been a growing appreciation of the role played by interactions in topologically non-trivial systems. In addition to changing the topological classification of systems \cite{yao2013}, interactions have been shown to modify topologically protected states, resulting in a range of exotic phenomena \cite{viyuela2018, neupert2011, sheng2011, beri2012, galpin2014, winkler2017}. The relative simplicity of one-dimensional topological superconductors (TSCs) with a finite charging energy \cite{fu2010, zazunov2011, zazunov2012, altland2014, plugge2015,bao2017} is particularly attractive for studying interaction effects in topological matter. A recent experimental investigation of this system\cite{albrecht2016} has offered compelling evidence for the existence of Majorana Bound States \cite{alicea2012} (MBSs) in condensed matter, whilst theoretical work indicates that it may host a delocalized many-body state arising from the interplay of interactions with MBSs \cite{vanbeek2017}. Previous studies have shown how a Josephson coupling in this system affects its conductance properties \cite{hutzen2012, didier2013}.

In this work, we exploit the topologically protected MBSs in the TSC to probe the non-equilibrium charge dynamics of a Josephson junction. The Josephson effect is one of the most prominent manifestations of superconducting phase coherence \cite{josephson1962}. Whilst the effect owes its existence to microscopic quantum objects (Cooper Pairs), at the macroscopic level it is essentially classical in nature. There are, however, other phenomena associated with superconductors that do not admit such a classical description. In particular, it was realized over thirty years ago \cite{likharev1985} that the competition between charging and Josephson energies in a Josephson junction results in a system whose behaviour is directly analogous to that of an electron in a periodic potential. Just as the electron's properties depend periodically on its momentum, with period given by the reciprocal lattice vector, the observables associated with the junction are $2e$ periodic in charge, where $e>0$ is the magnitude of the electronic charge. This periodicity is, fundamentally, contingent upon charge-phase conjugation and constitutes a macroscopic quantum phenomenon. That such a state of affairs can exist is interesting in its own right, and some experimental progress has been made in demonstrating that remarkable effects, such as Bloch Oscillations, can indeed be observed in such systems \cite{kuzmin1991, nguyen2007}. However, the $2e$ periodicity acts as a barrier to interrogation of the system, since for ideal superconductors all sub-gap charge perturbations must be in multiples of $2e$ and therefore do not change the state of the system.

By introducing a pair of MBSs into the system, we are not only able to overcome this obstacle, but also exploit the non-locality inherent to the MBSs. Taken together, the MBSs constitute a single fermionic state at zero energy which, due to interactions, persists even in rather short systems \cite{dominguez2017,escribano2018}. The MBSs therefore allow single electrons from an external reservoir to tunnel into and out of the system, thereby permitting perturbations of the junction's electronic state in a way that is qualitatively distinct from the Cooper pair processes considered previously. Furthermore, the delocalized nature of the fermionic state corresponding to the two MBSs means that it permits current flow over an extended distance through the TSC, in contrast to the sub-gap quasiparticles that have been considered previously \cite{buttiker1987, geigenmuller1988,mullen1988,fulton1989, vora2017}. As we will show, it is this current through the TSC that allows controlled sub-gap perturbations of the Josephson junction. We develop the theoretical formalism necessary to characterize such MBS mediated single-particle processes, and discuss the consequences of their existence on the charge dynamics of the Josephson junction. We demonstrate that the system exhibits a rich variety of dynamic regimes which can be explored experimentally by varying its electrical inputs.

In Section II we develop the theoretical framework necessary to describe the dynamics of the Majorana-Josephson system. We then analyze these dynamics and present our results in Section III before providing a summary of our conclusions in Section IV. Unless explicitly noted otherwise, the system parameters given in Appendix \ref{system parameters} are used throughout this work.

% ----------------------------------------------------------------------------

\section{Majorana-Josephson Hamiltonian}

\begin{figure}
	\centering
	\includegraphics[width=\columnwidth]{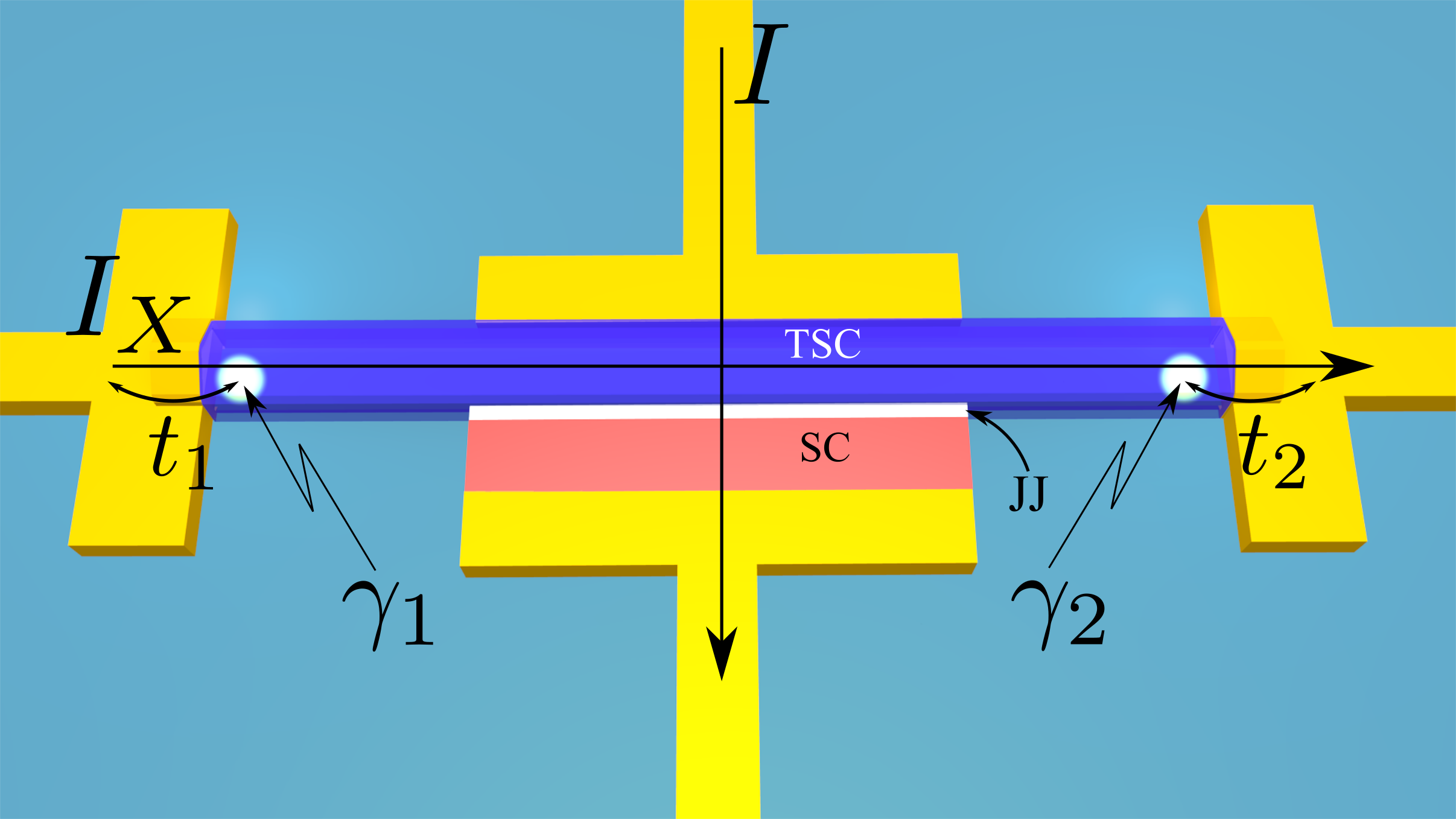}
	\caption{\label{setup}
		A floating topological superconductor (blue) hosting Majorana Bound States, $\gamma_{1,2}$ is coupled to normal metal leads (yellow) with tunnelling energies $\lambda_{1,2}$ and joined via an insulating weak link (white) to a grounded s-wave superconductor (red). A bias current $I$ is passed through the Josephson junction. A transverse current $I_X$ is established between the two metal leads, via the TSC, when there is a potential difference between them.
	}
\end{figure}

We consider the setup shown in Fig. \ref{setup}, in which a one dimensional floating topological superconductor which has MBSs at its ends  is coupled to three normal metal leads and connected, via a weak tunnelling junction, to a grounded s-wave superconductor. The behaviour of this system is the result of three distinct factors, namely charging energy, Josephson coupling and the MBSs. In this section we describe how we model these three components. The charging energy and Josephson coupling together give rise to quasicharge, which we discuss in subsections A and B, whilst the MBSs mediate a single-particle tunnelling process between the metallic leads and TSC, which is described in subsection C.

\subsection{Quasicharge and Band Structure}

\begin{figure}
	\centering
\includegraphics[width=\columnwidth]{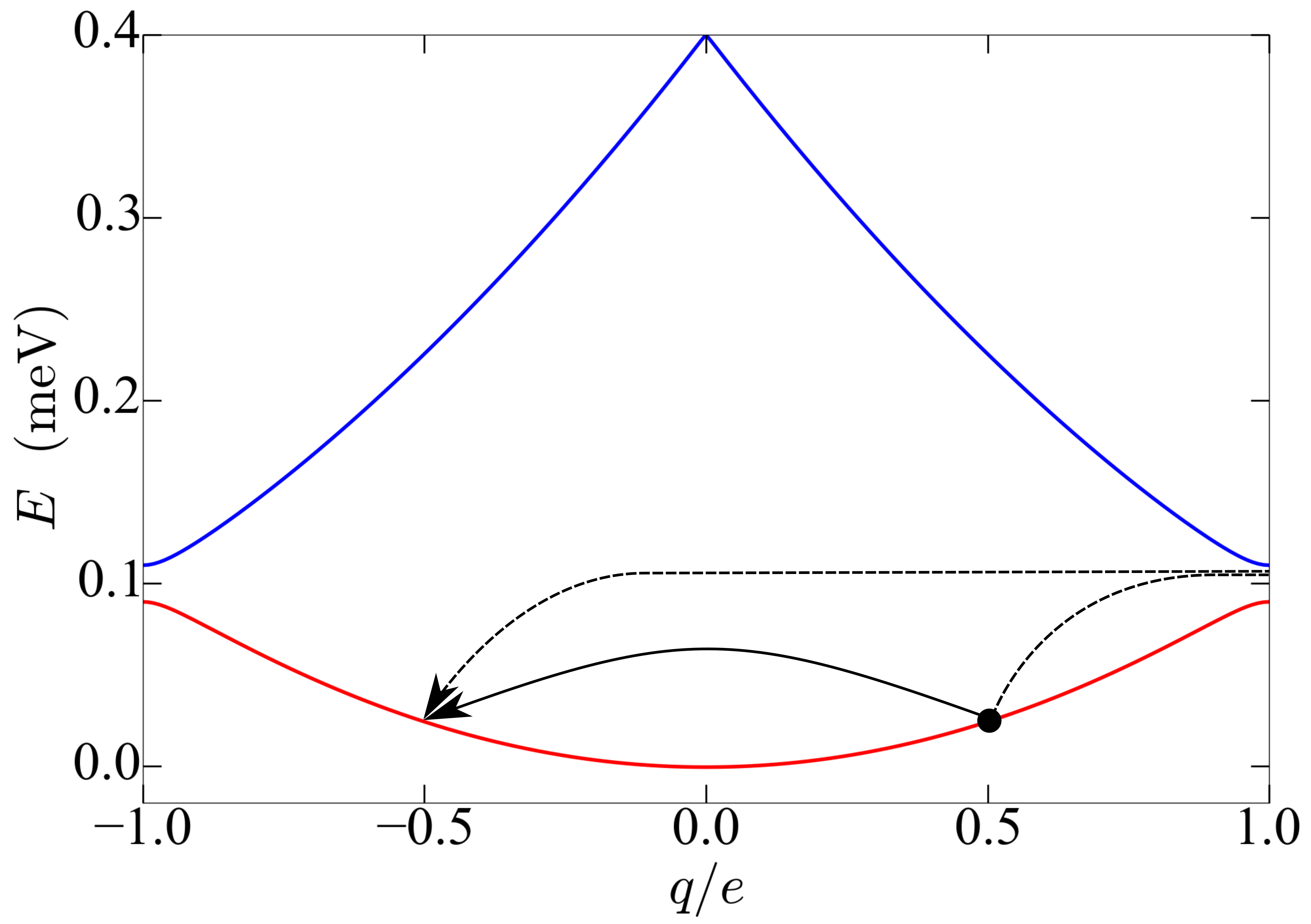}
\caption{\label{bandstructure}
	The band structure corresponding to Eq. \eqref{hsc} for $E_C = 0.1\text{meV}$ and $E_J = 0.02\text{meV}$. Only the first (red) and second (blue) bands are shown. Note that the bandwidth of the first band is $\sim E_C$ whilst the band gap between the two bands is $\approx E_J$. Also shown are two Majorana tunnelling events at a typical value of $q\approx 0.5e$. The solid black line represents tunnelling of an electron from the TSC to a metallic lead, thereby reducing $q$ by $e$. The dashed black line represents tunnelling of an electron from a metallic lead to the TSC, followed by a Bloch reflection in which a Cooper pair tunnels from the TSC to the s-wave superconductor, with the net result being that, once again, $q$ is reduced by $e$.
}
\end{figure}

A setup very similar to ours has previously been investigated in both theoretical \cite{fu2010,zazunov2011,zazunov2012,altland2014,plugge2015,vanbeek2017,hutzen2012} and experimental \cite{albrecht2016} work, with our study being distinguished by the addition of a current biased Josephson coupling. It is therefore straightforward to write down the Hamiltonian associated with the TSC\cite{likharev1985} (see also Ref. \onlinecite{buttiker1987}),
\begin{equation}\label{h0}
H_{sc} = \frac{Q^2}{2C}-E_J\cos\left(\phi\right),
\end{equation}
where $Q$ is the total charge difference across the Josephson junction between the TSC and s-wave superconductor, $C$ is the capacitance of the Josephson junction and $\phi$ is the phase of the TSC relative to the s-wave superconductor. As with all superconductors, the TSC obeys the charge-phase commutation relation,
\begin{equation}\label{qphicommutation}
\left[\phi,Q\right] = 2ei.
\end{equation}
Using this, we rewrite \eqref{h0} in terms of $\phi$ only,
\begin{equation}\label{hsc}
	H_{sc} =  -E_C\frac{\partial^2}{\partial\left(\phi/2\right)^2}-E_J\cos\left(\phi\right),
\end{equation}
where $E_C = \frac{e^2}{2C}$. Since the potential term in this Hamiltonian is periodic in $\phi$, the solutions will take the familiar, periodic, Bloch form. In particular, the energies of the Hamiltonian are given by $E_{s}\left(q\right)$ where $s$ is a band index and $E_s\left(q\right) = E_s \left(q+2e\right)$. The quasicharge, $q$, is directly analogous to the quasimomentum in a crystal lattice. It corresponds to the total charge on the TSC, modulo $2e$. The first two energy bands of Eq. \eqref{hsc} are shown as a function of $q$ in Fig. \ref{bandstructure}. Throughout the remainder of this work we will assume that the system is always in the lowest energy band and neglect inter-band processes. More details regarding the justification for, and consequences of, this assumption can be found in Appendix \ref{inter-band transitions}.

\subsection{Slow Quasicharge Evolution}

\noindent In addition to the charging and Josephson energies of the superconductor, the total Hamiltonian of the system also includes the current-phase interactions \cite{larkin1984},
\begin{equation}
	V_I = -\frac{\hbar}{2e}I\left(t\right)\phi, \hspace{5pt} V_q =\frac{\hbar}{2e}I_q\phi,
\end{equation}
in which $I(t)$ is the (possibly time dependent) bias current applied to the junction and $I_q$ is a leakage current arising from the voltage across the Josephson junction associated with charge accumulation and carried via sub-gap quasiparticles in the superconductor, which exist independently of the MBSs. The exact origin of these quasiparticles is uncertain, indeed, they may have multiple sources, with the dominant source depending on the sample in question, but the existence of the quasiparticle current is an empirical fact \cite{martinis2009} and so we include  it in our model without overly concerning ourselves with its microscopic origin. Substituting  $E_{s}\left(q\right)$ for the contribution to the Hamiltonian that comes solely from the superconductor, we see that the total junction Hamiltonian is given by,
\begin{equation}\label{junctionhamiltonian}
H_{JJ} = E^{(s)}\left(q\right) -\frac{\hbar}{2e}I\left(t\right)\phi + \frac{\hbar}{2e}I_q\phi.
\end{equation}
As the first term in Eq. \eqref{junctionhamiltonian} depends only on $q$ it is clear, from the commutation relation Eq. \eqref{qphicommutation}, that the time evolution of $q$ depends only on the phase-current interaction terms and is given by, $\dot{q} = I(t) - I_q$. The quasiparticle current, $I_q$, is written formally as the product of quasiparticle mediated conductance, $G\left(\omega\right)$, and voltage, $V$, across the junction, $I_q = G(\omega)V$. In the single band approximation $V$ is simply equal to $\text{d}E_0/\text{d}q$. Furthermore, the quasiparticle conductance is a constant $G(\omega)=G$, provided\cite{likharev1985} $\omega \ll \frac{\Delta}{\hbar}$. Typically, $\Delta \approx 0.1\text{meV} $ and so $G$ is constant for $\omega \ll 10^{11}\text{s}^{-1}$, which is true throughout the range of driving frequencies we study. Nevertheless, since $G$ is a function of quasiparticle density, the exact value of $G$ will vary depending on the superconductor and its environment \cite{sun2012}. Whilst this does introduce a random component to the value of $G$, and by extension $I_q$, previous work indicates that, for any given sample, $G$ may be treated as constant over the timescales considered in this paper \cite{martinis2009,vora2017}. We therefore arrive at a straightforward Langevin-type equation for the quasicharge,
\begin{equation}\label{qevolution}
	\dot{q} = I(t) - G\frac{\text{d}E_0}{\text{d}q}.
\end{equation}

\begin{figure}
	\centering
	\includegraphics[width=\columnwidth]{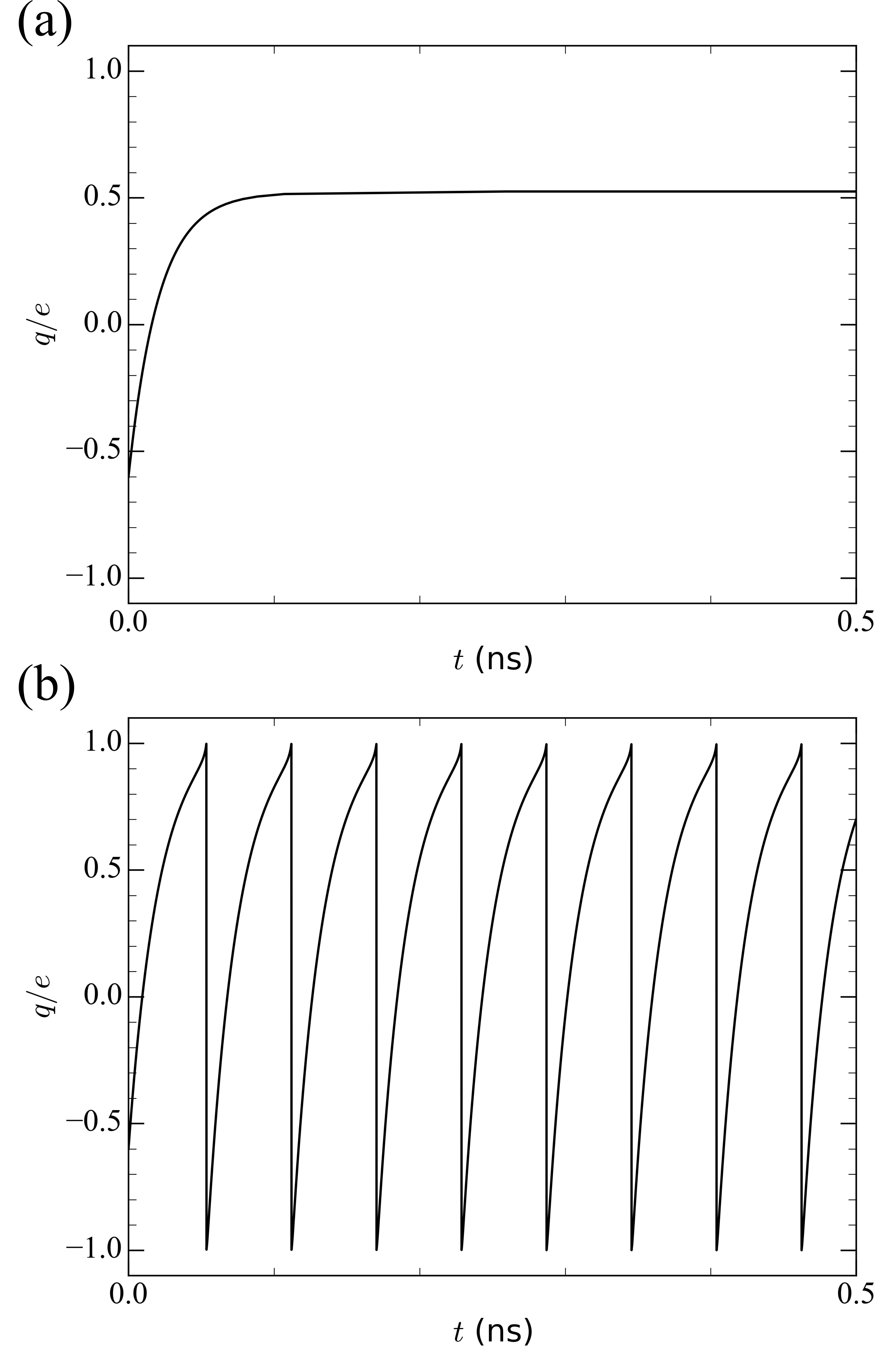}
	\caption{\label{qevobasic}
		Evolution of quasicharge with time in the case of no Majorana tunnelling, with initial quasicharge $q_0 = -0.6e$. (a) A bias current of $4.0\text{nA}$ results in the quasicharge tending to a fixed value, $q=0.52e$. (b) A bias current of $8.0\text{nA}$ gives rise to Bloch oscillations, as expected for $I_B = 6.2\text{nA}$.
	}
\end{figure}

\noindent This evolution of the quasicharge is a result of both the bias current and the band structure resulting from the charging and Josephson energies. By analyzing Eq. \eqref{qevolution} we conclude that the system exhibits two regimes. For low currents, specifically,
\begin{equation}\label{thresholdcurrent}
\frac{I}{G} <\text{max}\left(\frac{\text{d}E_0}{\text{d}q}\right),
\end{equation}
the quasicharge tends to a fixed point, $q_0$ where,
\begin{equation}\label{fixedpointvalue}
\left.\frac{\text{d}E_0}{\text{d}q}\right|_{q_0}=\frac{I}{G}.
\end{equation}
Whilst for currents greater than those in Eq. \eqref{thresholdcurrent} the quasicharge never assumes a constant value. From Eq. \eqref{qevolution}, $\dot{q} > 0$ at all times and so, since $q$ is only defined modulo $2e$, the system executes Bloch oscillations with period,
\begin{equation}
\tau_B = \int_{-e}^{+e}\frac{\text{d}q}{I-G\frac{\text{d}E}{\text{d}q}}.
\end{equation}
We therefore define the Bloch oscillation threshold current, $I_B = G\text{max}\left(\frac{\text{d}E_0}{\text{d}q}\right)$. Physically, these Bloch oscillations correspond to tunnelling of a Cooper Pair across the Josephson junction. These two cases, a static quasicharge for bias currents given by Eq. \eqref{thresholdcurrent} and Bloch oscillations at larger currents,  are illustrated in Fig. \ref{qevobasic} (a) and (b) respectively, in the absence of Majorana tunnelling.

\subsection{Majorana-Mediated Single Particle Tunnelling}

The results described above are a generic feature of Josephson junctions with a charging energy and do not depend upon the presence of MBSs. However, by considering the setup in Fig. \ref{setup} we find that the system has the potential to exhibit a much wider range of interesting phenomena when accompanied by MBSs. The presence of MBSs is notable, not only because they offer the possibility of single-particle tunnelling into the floating superconductor, below the superconducting gap, but also because their non-local nature enables transmission of current across the TSC. To determine the effects of this process, we begin by finding the tunnelling rates associated with the MBS. The Hamiltonian describing tunnelling between the normal metallic leads and superconductor has been found previously by projecting the operators of the electrons in the superconductor onto an MBS manifold\cite{flensberg2010} and is given by,
\begin{equation}\label{tunnelham}
H_T = \sum_{j,k} \lambda_j c^\dagger_{j,k}\gamma_je^{-\frac{i\phi}{2}}+\text{h.c.},
\end{equation}
where $j=1,2$ indexes the two leads, $\lambda_j$ are the tunnelling energies, $c_{j,k}$ is the operator for a fermion in lead $j$ with momentum $k$ and $\gamma_j$ are the Majorana operators. The operator $e^{-\frac{i\phi}{2}}$ corresponds to annihilation of an electron in the superconductor and is required to ensure charge conservation. This factor, in concert with the charge conserving representation of $\gamma$ gives rise to normal and anomalous tunnelling \cite{zazunov2011,vanbeek2017,hutzen2012}. Note that we assume negligible overlap of MBS $\gamma_{1(2)}$ with lead $2(1)$, which is valid provided that the TSC is much longer than its coherence length. Even if this length condition were not true, a small overlap of MBS $\gamma_{1(2)}$ with lead $2(1)$ would not significantly affect our results. Furthermore, self-interaction effects of MBSs work against the energy splitting by the overlap and can cause a further pinning of the MBSs to zero energy.\cite{dominguez2017,escribano2018}
The absence of spin degeneracy in Eq. \eqref{tunnelham} is due to the spin polarisation of the MBSs \cite{sticlet2012}, allowing electrons to be treated as spinless fermions for the purposes of tunnelling, despite the lead electrons being spinful. The spin polarization is inessential to the results reported in this paper, but nonetheless is a feature of MBSs and may have some relevance in the case of coupling to a different type of system, such as a ferromagnetic lead. A straightforward application of Fermi's Golden Rule yields the tunnelling rate corresponding to $H_T$,
\begin{equation}\label{tunnelrate}
	\Gamma_{MBS} = \Gamma_1\zeta\left(\delta E_{ch},V_1\right)+\Gamma_2\zeta\left(\delta E_{ch},V_2\right),
\end{equation}
where $\Gamma_j = \rho\lambda_j^2/\hbar$, with $\rho$ the density of states in the metallic leads, and $\zeta$ is a combination of particle and hole Fermi functions given by,
\begin{equation}\label{zeta}
\zeta\left(\delta E_{ch}, V\right)  = \frac{1}{e^{\frac{\delta E_{ch}+eV}{k_BT}}+1}+\frac{1}{e^{\frac{\delta E_{ch}-eV}{k_BT}}+1},
\end{equation}
where $T$ is the electron temperature and $k_B$ is the Boltzmann constant. We have assumed that the density of states is identical in both the left (1) and right (2) leads, but that each lead has a voltage bias $V_{1,2}$. By $\delta E_{ch}(q)$ we denote the (quasicharge dependent) energy change on tunnelling of a single particle into or out of the TSC from the leads. We note that the dependence of $\delta E_{ch}$ on quasicharge alone, and not whether tunnelling is to or from the TSC, is a direct consequence of the particle-hole symmetry imposed on the system by the Josephson coupling. To be more specific, the $2e$ periodicity in quasicharge mentioned above means that tunnelling of either a particle or hole from a lead into the TSC results in the same energy change $\delta E_{ch}\left(q\right)$ in both cases, for any $q$. For an island without this $2e$ periodicity, particle and hole tunnelling events are inequivalent and consequently have different charging energies associated with them, which results in a more complicated form for $\Gamma_{MBS}$. However, as can be seen from Eq. \eqref{zeta}, the inherent particle-hole symmetry of Eq. \eqref{tunnelrate} is broken by a finite bias voltage, $V_{1,2}$.

The impact of Eq. \eqref{tunnelrate} on the charge dynamics of the Majorana-Josephson device can be summarized as follows. At low temperatures, $k_BT \ll |\delta E_{ch}\pm eV|$, we see from Eq. \eqref{zeta} that $\zeta \simeq 0$ when both $\delta E_{ch} + eV > 0$ and $\delta E_{ch}- eV > 0 $, whilst if $\delta E_{ch} + eV < 0$ or $\delta E_{ch} - eV < 0$, or both expressions are less than zero, then $\zeta$ is of order $1$ and tunnelling is likely. Since the factor $\Gamma_{1,2}$ in Eq. \eqref{tunnelrate} is typically very large, the above observation implies that, in the low temperature limit, $\Gamma_{MBS}$ transitions rapidly from zero to some very large number, as the values of $\delta E_{ch}$ and $eV$ change. From the expression for the charging energy, $\delta E_{ch}$, we find that, in the $T = 0$ limit, the tunnelling rate $\Gamma_{MBS}$ is zero for $|q| < \frac{e}{2}\left(1-\frac{V}{E_C}\right)$ and very large otherwise. At finite temperatures the step boundary between tunnelling and non-tunnelling regimes is softened, but nonetheless we can identify an absolute value of the quasicharge above which tunnelling proceeds at a rapid rate and below which tunnelling is very slow. In particular, as the applied voltages, $V_{1,2}$, tend to zero, the threshold value of the quasicharge tends to $|q| = e/2$.
\newline

In addition to the MBSs there could be, in principle, other sub-gap quasiparticle states in the TSC \cite{guinea86}, which may originate from thermal excitations or unintentional electromagnetic irradiation \cite{aumentado2004}. Previous experimental studies on superconducting qubits \cite{martinis2009,sun2012} have found that the single-particle tunnelling rate corresponding to these quasiparticles is, $\Gamma_{QP} \sim 10^6 \text{s}^{-1}$ which is much less than the typical rate associated with the MBSs, $\Gamma_{MBS} \sim 10^{11} \text{s}^{-1}$, and so we safely neglect the influence of these non-topological quasiparticles. It is worth noting that, even if $\Gamma_{QP}$ and $\Gamma_{MBS}$ were comparable, the presence of the MBSs would give rise to qualitatively different effects from the quasiparticles. This is due to the well defined energy of the MBSs, compared with the continuum of energies adopted by the quasiparticles, which results in $\Gamma_{MBS}$ being proportional to a Fermi function, whilst $\Gamma_{QP}$ is proportional to a Bose function and so the two rates have qualitatively different temperature, $E_C$ and $V_{1,2}$ dependence. Furthermore, the delocalized nature of the single particle state associated with the MBSs enables charge transport that would not necessarily be possible in the presence of non-topological quasiparticles alone.
\newline

A final point to consider in regard to tunnelling between the TSC and metallic leads is the influence of memory effects. That is to say, the impact of a given tunnelling event on the probability of subsequent tunnelling events taking place. The most significant effect is that tunnelling changes the total charge on the TSC island, and the influence this has on future tunnelling probabilities is captured in the $\delta E_{ch}$ terms that appear in Eq. \eqref{zeta}. In principle there is an additional process which should be considered, in which the tunnelling event modifies the quantum state of the TSC beyond simply changing the total number of electrons. In this work we do not take into account the impact of this second consideration, for two reasons: firstly, the change in tunnelling probability associated with this process is likely to be negligible compared to the influence of macroscopic charging effects; secondly, a previous study into the relaxation of charge excitation ``hotspots" in current biased superconductors \cite{marsili2016} found that the system typically relaxed after around $50$ps, which is shorter than the time scale of almost all the processes we describe here. This figure, $50$ps, is likely to be much longer than the time scale of the processes we are neglecting, since it relates to an essentially classical excitation, less susceptible to environmental damping. Nevertheless, it is possible that the very fast ``ringing" phenomena which we describe later will be modified by quantum memory effects and this would be an interesting effect to study theoretically or experimentally in the future.
% ----------------------------------------------------------------------------
\section{Device Dynamics}

\begin{figure}
	\begin{minipage}[t]{\columnwidth}
		\centering
		\includegraphics[width=\columnwidth]{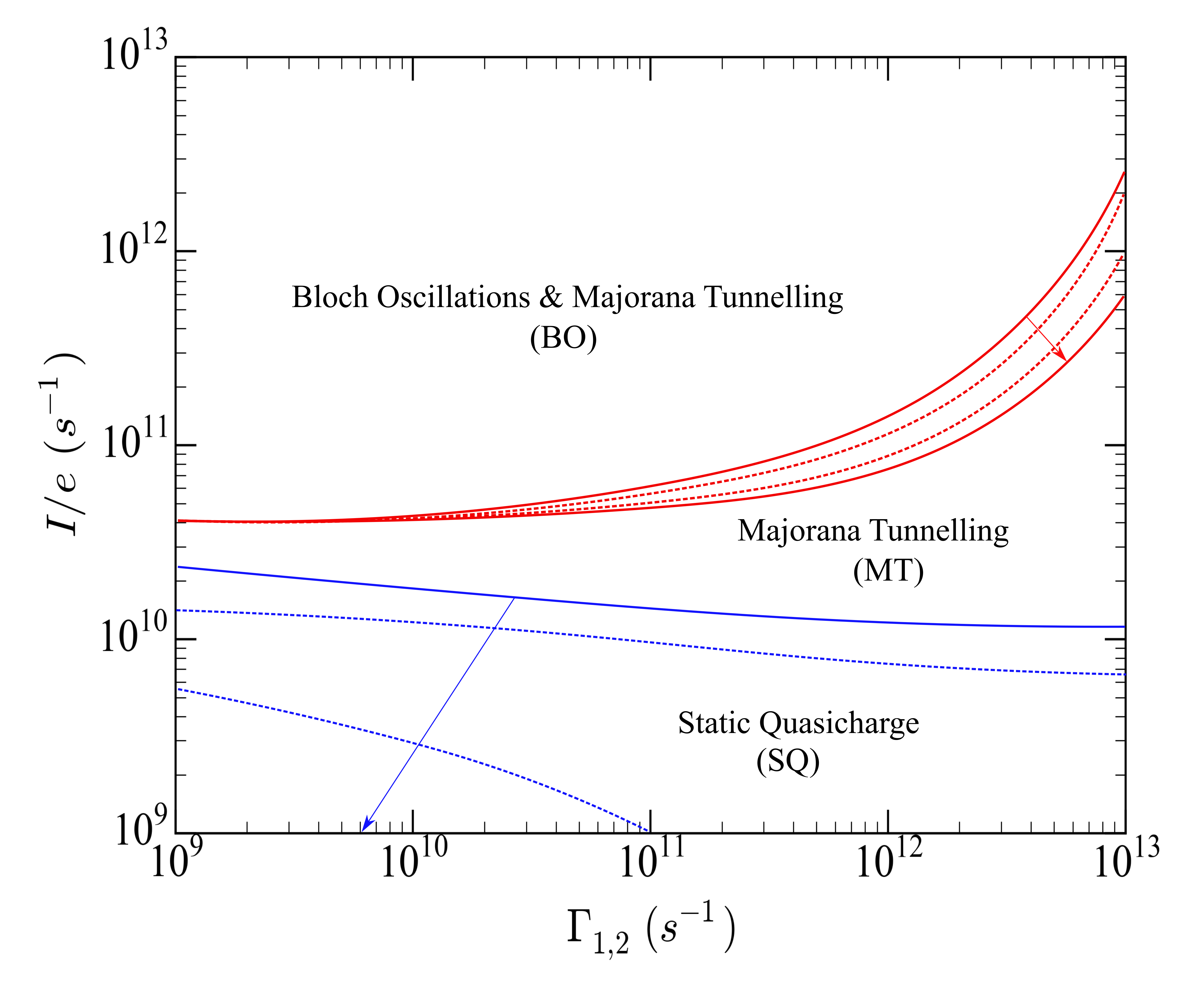}
		\caption{\label{ICregimediagram}
			Regime diagram for the Majorana-Josephson system, plotted in terms of bias current, $I$, across Josephson junction and tunnelling rate $\Gamma_{1,2}=\Gamma_1=\Gamma_2$ from normal leads to TSC. SQ-MT and MT-BO regime boundaries are shown in blue and red, respectively. Solid lines indicate regime boundaries for bias voltages $V_2 = -V_1 = 0.0$ or $0.1 \text{mV}$, whilst dashed lines correspond to bias voltages of $V_2 = -V_1 = 0.03$ or $0.07 \text{mV}$, with arrows indicating increasing voltage magnitude. Note that at a bias voltage of $0.1 \text{mV}$ the SQ regime is extinguished and no SQ-MT boundary is visible.}
	\end{minipage}
\end{figure}

Several parameters influence the behaviour of the Majorana-Josephson system, such that it is impractical to simultaneously capture the effect of all of them in a single analysis. However, in the case of a static bias current, there are three main quantities of interest, namely the magnitude of the bias current, $I$ that appears in Eq. \eqref{qevolution}, the tunnelling rates from the normal leads to the TSC, $\Gamma_{1,2}$, and the bias voltages, $V_{1,2}$, of the leads. By considering only the impact of variations in these three quantities, it is possible to describe the salient features of the Majorana-Josephson system's dynamics in an easily accessible manner.

We determine the dynamics of the Majorana-Josephson system by solving Eq. \eqref{qevolution} with the classical Runge-Kutta method and incorporate the influence of the MBSs by using a Monte Carlo approach to find the tunnelling rate given by Eq. \eqref{tunnelrate}. Full details of this procedure can be found in Appendix \ref{numerical method}.
\subsection{Time Evolution of Quasicharge}
Quasicharge is the most basic quantity upon which other dynamic variables depend, and so we begin by establishing a comprehensive picture of quasicharge dynamics throughout the whole of the system's parameter space. This information is presented in the regime diagram shown in Fig. \ref{ICregimediagram}.

\begin{figure}
		\centering
		\includegraphics[width=\columnwidth]{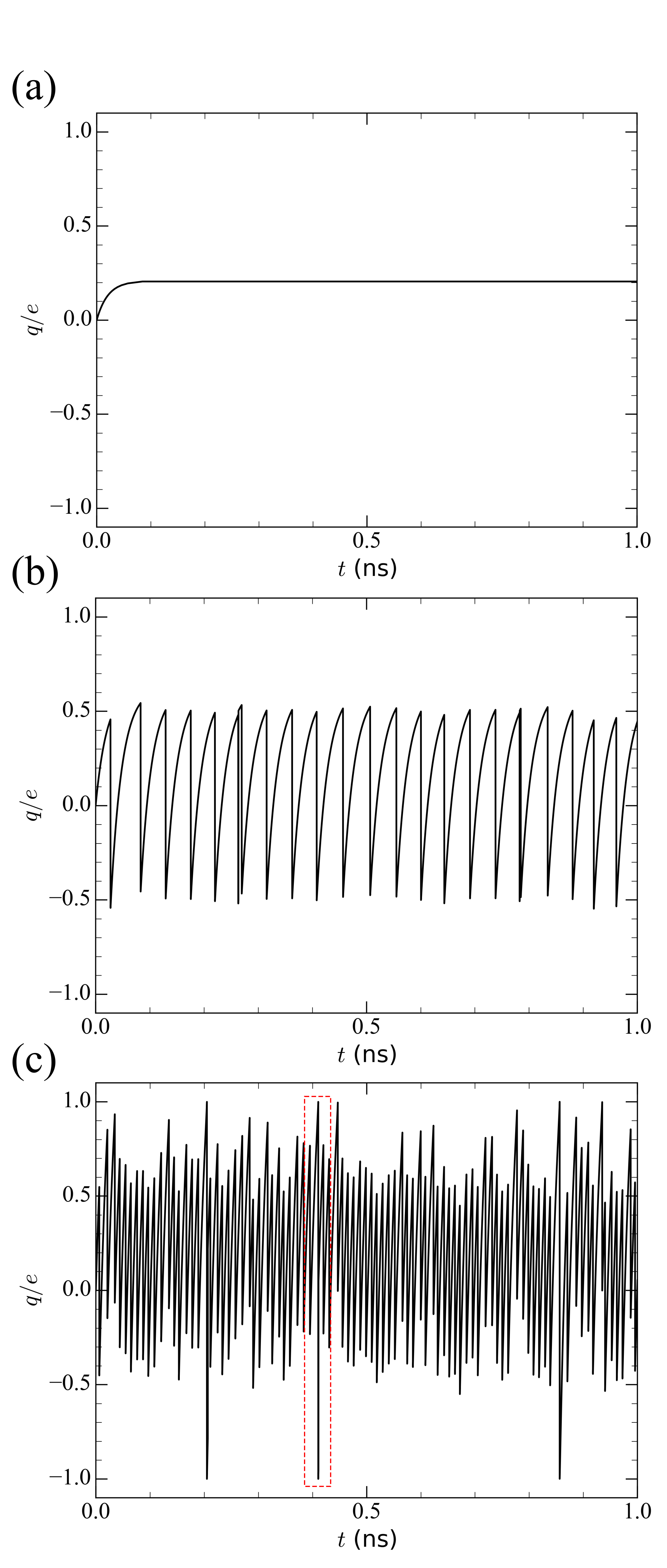}\caption{\label{regimes} Examples of quasicharge behaviour for the three different regimes shown in Fig. \ref{ICregimediagram}. (a) Static Quasicharge, SQ ($I=1.6\text{nA}$). (b) Majorana Tunnelling, MT ($I=4.8\text{nA}$). (c) Bloch Oscillations, BO ($I=16.0\text{nA}$). The red dashed outlines in (c) indicate Bloch Oscillations.}
\end{figure}

We sort the behaviour of the system into three broad categories: Static Quasicharge (SQ), for which the bias and leakage currents in Eq. \eqref{qevolution} exactly balance and the quasicharge remains at a constant value below $0.5e$; Majorana Tunnelling (MT), where the bias current, $I$, is not sufficiently large to drive the quasicharge to the zone boundary, but nonetheless is large enough to force the system into a regime where MBS mediated tunnelling becomes appreciable; Bloch Oscillations and Majorana Tunnelling (BO), in which tunnelling rates are appreciable, as in MT, but $I$ is sufficiently large to drive the quasicharge to the zone boundary, resulting in Bloch oscillations. Note that whilst we denote this regime simply BO for convenience, the dynamics of the system consists primarily of Majorana tunnelling, with occasional Bloch oscillations. Examples of the different regimes are shown in Fig. \ref{regimes}. We note, in particular, the difference between Fig. \ref{regimes}c and Fig. \ref{qevobasic}b, which highlights the effect of the MBSs, namely enabling single particle tunnelling and consequently suppressing Bloch oscillations. 

We point out that choosing to classify the Majorana-Josephson system according to these three regimes is somewhat arbitrary, particularly in the case of BO since there is little meaningful physical distinction between Bloch oscillations resulting from slow evolution of the quasicharge to the zone boundary, as in BO, and those Bloch reflections caused by Majorana tunnelling events that rapidly drive the system outside of the quasicharge Brillouin Zone, as occurs in both BO and MT. Furthermore, the stochastic nature of the system behaviour means that the position, and indeed existence, of the regime boundaries in Fig. \ref{ICregimediagram} is not a universal property, but rather depends on the timescale over which the system is studied. In the long time limit, the SQ regime no longer exists and the MT-BO boundary is a line of constant $I$. Nevertheless, we contend that the classification shown in Fig. \ref{ICregimediagram} is meaningful, in that the behaviour of the system does change significantly as its parameters change, but we caution against interpreting Fig. \ref{ICregimediagram} as a phase diagram in the usual sense of the term.
\newline

It is straightforward to understand the general form of Fig. \ref{ICregimediagram}. The bias current sets the long-time, zero-tunnelling, equilibrium quasicharge, in accordance with Eq. \eqref{fixedpointvalue}. There are essentially three distinct bias current ranges: when the bias current is less than the threshold current for Majorana tunnelling, $I_\theta$, we have $I<I_{\theta} = G\left.\frac{\text{d}E_0}{\text{d}q}\right|_{e/2}$ and the system tends to a steady state with $q<e/2$; when the bias current is greater than $I_\theta$, but less than the Bloch oscillation threshold current, $I_B$, we have $I_{\theta}<I<I_B = G\text{max}\left(\frac{\text{d}E_0}{\text{d}q}\right)$ and the equilibrium quasicharge is in the range $e/2<q<e$; at large bias currents, $I>I_B$, the system does not adopt a stable value of $q$ but rather, in the zero-tunnelling limit, executes Bloch Oscillations.

Since the probability of MBS mediated tunnelling becomes very large for $q>0.5e$ (if $T\simeq 0, V_{1,2}=0$), for the system to be in the SQ regime, it is necessary that $I<I_{\theta}$, which is supported by Fig. \ref{ICregimediagram}. However, for high tunnelling rates, $\Gamma_{1,2}$, even at $q\lesssim 0.5e$ the probability of tunnelling can be appreciable and so the SQ regime persists only to lower values of bias current, as can be seen in Fig. \ref{ICregimediagram}.

Similarly, for $I<I_B$ there is no possibility of Bloch Oscillations, which is consistent with the observation that the MT-BO regime boundary does not descend below $I_B$ in Fig. \ref{ICregimediagram}. We also see that, as the tunnelling rate increases, the MT-BO boundary shifts linearly to higher bias currents. In essence, an increase in the Majorana tunnelling rate decreases the probability that the quasicharge will evolve slowly to the zone boundary without undergoing a discrete jump due to Majorana tunnelling. A larger bias current is therefore required to more quickly drive the quasicharge towards the zone boundary. In the next section we shall discuss in more detail the role that Majorana tunnelling has to play in the promotion or suppression of Bloch oscillations.

\subsection{Bias Voltage Dependence}
Figure \ref{ICregimediagram} also shows how the regime boundaries evolve on changing the bias voltages, $V_{1,2}$ in the left and right normal leads. The red and blue arrows indicate increasing bias voltage magnitude. We see that the SQ-MT boundary shifts to progressively lower values of bias current as $|V_{1,2}|$ increases. This is explained by examining the role of bias voltage in Eq. \eqref{zeta}. For $V=0$ and $T\simeq0$, the exponential term in the denominator of $\zeta$ is large for $\delta E_{ch} > 0 $ and so the tunnelling rate is small for values of $q$ corresponding to $\delta E_{ch} > 0$, viz. $q < e/2$. However, if $V \neq 0$, then even when  $\delta E_{ch} > 0$ one of the two exponentials in Eq. \eqref{zeta} will be small, provided $\delta E_{ch} \pm eV < 0$ in which case the tunnelling rate will be large despite the charging energy associated with tunnelling being positive. As $|V|$ increases, progressively more positive values of $\delta E_{ch}$ conform to the requirement $\delta E_{ch} \pm eV < 0$ and so the region in $q$-space where tunnelling rates are appreciable grows. That is to say, if $V = 0$ tunnelling is only appreciable for $|q| > e/2$, but if $V \neq 0 $, then tunnelling is appreciable for $|q| > \frac{e}{2}\left(1-\frac{V}{E_C}\right)$.  The bias current, $I$, determines the equilibrium value of the quasicharge according to Eq. \eqref{fixedpointvalue} with lower $I$ corresponding to lower values of $q_0$. Consequently, as $|V|$ grows, increasing the range of quasicharge values for which tunnelling is appreciable, the SQ region, where tunnelling is negligible, corresponds to progressively lower values of the bias current.

The movement of the MT-BO regime boundary is, at first, more surprising. We previously discussed how, at high tunnelling rates, Majorana tunnelling leads to suppression of the BO region. We have also just seen how increasing bias voltage results in Majorana tunnelling in more of the quasicharge space. We might, therefore, expect increasing bias voltage to suppress the BO regime, but from Fig. \ref{ICregimediagram} we see that the opposite is true: as bias voltage increases, the BO regime grows. To understand this result, we must fully appreciate the role that Majorana tunnelling plays in inhibiting or promoting Bloch oscillations. For a Bloch oscillation to take place, the quasicharge must evolve slowly to the zone boundary (as distinct from a Bloch reflection which occurs whenever the quasicharge reaches the zone boundary, slowly or by a sudden jump). Any processes which take the quasicharge closer to the zone boundaries therefore promote Bloch oscillations, whilst those that take $q$ further from the zone boundaries inhibit Bloch oscillations. If tunnelling of a particle or hole takes place when $|q|>0.5e$ then $|q|$ decreases, whilst if tunnelling takes place for $|q|<0.5e$, $|q|$ increases, i.e. moves closer to a zone boundary. It follows that any change in the system parameters that increases the Majorana tunnelling rate for $|q|>0.5e$ will decrease the probability of a Bloch Oscillation occurring, whilst changes that increase the tunnelling rate for $|q|<0.5e$ will increase this probability. Recalling the preceding discussion on the SQ-MT boundary's movement with increasing bias voltage, we see that non-zero $V_{1,2}$ increases the total tunnelling rate for $|q|<0.5e$ whilst having only a negligible impact for $|q|>0.5e$, with the effect becoming more pronounced at larger $|V_{1,2,}|$. We therefore anticipate that the BO region will grow as $|V_{1,2}|$ increases, which we see in Fig. \ref{ICregimediagram} is indeed the case.

\subsection{Transverse Current Switching}
\begin{figure}
		\centering
		\includegraphics[width=\columnwidth]{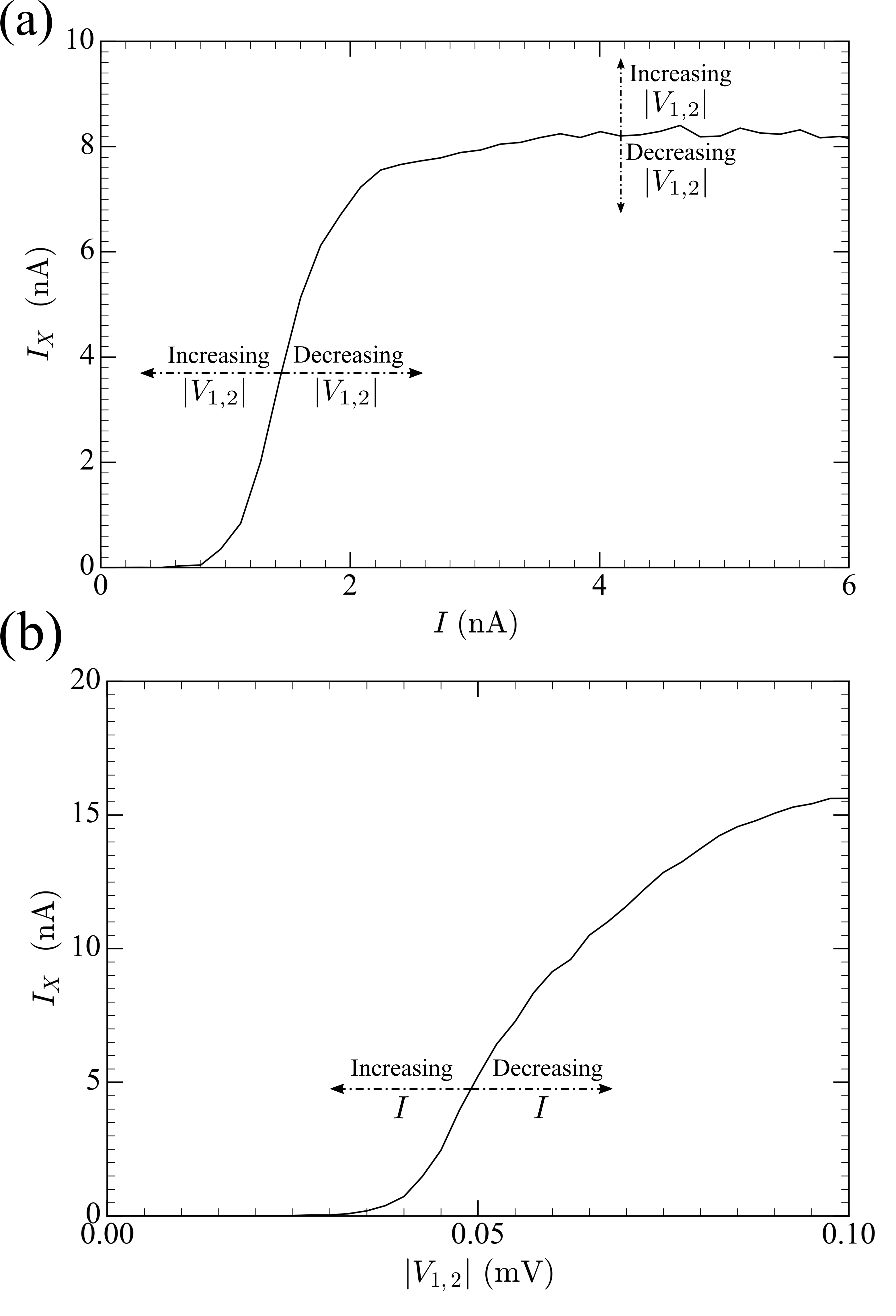}\caption{\label{electrics} Electrical properties of the Majorana-Josephson device. (a)  Time averaged $I_X$ vs bias current, $I$, across the Josephson junction. (b) Time averaged transverse current $I_X$ vs bias voltage, $V_2 = -V_1$, between the leads and TSC. Arrows indicate the qualitative change in (a) and (b) on changing $|V_{1,2}|$ and $I$, respectively. In (a), $V_2=-V_1=0.05\text{mV}$ whilst in (b) $I = 1.6\text{nA}$.}
\end{figure}

We now consider the electrical properties of the Majorana-Josephson device, as shown in Fig. \ref{electrics}. Considering the transverse current, $I_X$, that is transmitted across the TSC between the normal leads biased at $V_{1,2}$, the system acts as a transistor controlled either by the bias current, $I$, across the Josephson junction, or the bias voltage, $V_{1,2}$, across the TSC. Referring back to Fig. \ref{ICregimediagram}, $I_X = 0 $ when the system is in the SQ regime: no tunnelling implies no transfer of charge from the leads to the TSC and therefore no transverse current. In both the MT and BO regimes, tunnelling takes place at a high rate, resulting in an appreciable current. We note that our analysis includes only first order sequential tunnelling processes, an approximation valid in the large $E_C$ regime where second order tunnelling processes are strongly suppressed. Since the system is not gated to a charge degeneracy point \cite{hutzen2012}, but rather achieves charge degeneracy only intermittently due to the accumulation of charge caused by the bias current, $I$, the zero bias peak that is often regarded as a key characteristic of the MBSs does not contribute in a special way to $I_X$. Instead of remaining at the charge degeneracy point, the system is immediately driven away to different charging values.

Fixing the bias current and changing $V_{1,2}$ causes the SQ-MT regime boundary of the device to shift, as depicted in Fig. \ref{ICregimediagram}. Provided that the bias current and tunnelling rates are sufficiently low (such that the SQ regime is accessible in the first place) the system will cross the SQ-MT regime boundary at some finite bias voltage and transition from an insulating to conducting state, as shown in Fig. \ref{electrics}(b). The exact voltage at which this occurs depends linearly on $I$ and exhibits a non-linear dependence on the tunnelling rate from leads to TSC. Similarly, if the bias voltage is held at a sufficiently low value for the SQ regime to have a finite size, and the bias current is increased, the system will cross the SQ-MT phase boundary and go from the insulating to conducting state. This scenario is shown in Fig. \ref{electrics}(a). The bias current at which the system switches from an insulating to a conducting state depends linearly on $V_{1,2}$ and has a non-linear dependence on the MBS tunnelling rate. From Fig. \ref{ICregimediagram} we can see that, in general, the regime occupied by the Majorana-Josephson system has a rather weak dependence on the MBS-mediated tunnelling rate, compared to the stronger dependence on $V_{1,2}$ and $I$. 

\subsection{Time Dependent Driving Currents}

Thus far, we have concerned ourselves only with static driving currents, but we now consider the effects of applying a time-varying bias current, $I = I(t)$. In particular, we imagine a current of the form $I = I_{DC} + I_{AC}\cos\left(2\pi f t\right)$, with $I_{DC}, I_{AC} > 0$ and study the response of the Majorana-Josephson system over a range of current amplitudes and frequencies.

\begin{figure}
	\centering
	\includegraphics[width=\columnwidth]{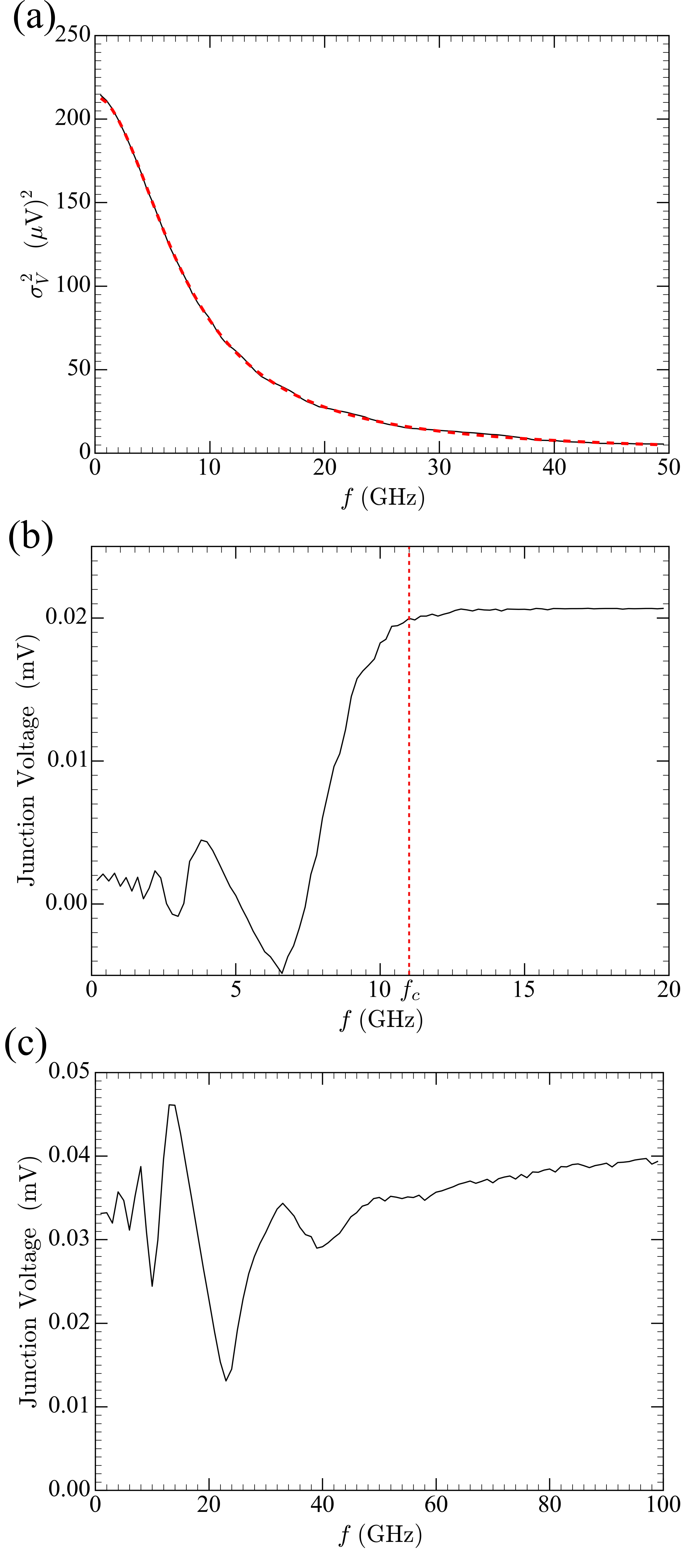}\caption{\label{VvsF}Time averaged voltage across Josephson junction, or variance of this voltage, as a function of bias current frequency for three different regimes. In all cases $I_\theta = 3.8 \text{nA}$. (a) $I_{DC}, I_{AC} = 0.8\text{nA}\ll I_{\theta}$ and so $V(f)$ is approximated by Eq. \eqref{approxV}, meaning that $\left<V\right>_t\simeq I_{DC}/G$. We therefore plot the variance of $V$ (solid black line) and compare it with the expected analytic result (dashed red line). (b) $I_{DC} = 0.8\text{nA}\ll I_{\theta}, I_{AC} +I_{DC} = 4.8\text{nA} \gtrsim I_{\theta}$ and $\left<V\right>_t$ is suppressed below some cut-off frequency, $f_c$, marked by a dashed red line, whilst adopting a fixed value above it. (c) $I_{DC} = 4.0\text{nA} \gtrsim I_{\theta}, I_{AC} = 4.0\text{nA} \neq 0$ and the average junction voltage exhibits resonances at low frequencies, before increasing to a constant value at higher frequencies.}
\end{figure}

There are two driving frequency-dependent quantities of interest: the voltage across the Josephson junction, $V = \frac{\text{d}E_0}{\text{d}q}$, and the transverse current through the TSC, $I_X$. We note that whilst the presence of a frequency-dependent junction voltage is a generic feature of any capacitive Josephson junction \cite{vora2017}, the existence of a transverse current $I_X$ is contingent upon the sub-gap Majorana bound states.

By considering the magnitudes of $I_{DC}$ and $I_{AC}$ relative to the threshold current $I_{\theta}$, we identify three different regimes of interest, namely: the low bias regime, $I_{DC}, I_{AC} \ll I_{\theta}$; the intermediate bias regime, $I_{DC}\ll I_{\theta}, I_{DC}+I_{AC} \gtrsim I_{\theta} $; and the high bias regime, $I_{DC}\gtrsim I_{\theta}, I_{AC} \neq 0$. These three regimes originate from the behaviour of $q$ with varying driving frequency. If driving is in the low current regime, $I_{DC}, I_{AC} \ll I_{\theta}$, then $I(t)<I_{\theta}$ for all $t$ and so $q$ never reaches a large enough value for Majorana tunnelling to be significant. In the intermediate current regime, $I_{DC}\ll I_{\theta}, I_{DC}+I_{AC} \gtrsim I_{\theta} $, we see that $I(t) \lessgtr I_{\theta}$, depending on the value of $t$. We might therefore expect Majorana tunnelling to take place at some point over one period of the bias current. However, this is not the case at high frequencies where, even though $I(t) > I_{\theta}$ for some values of $t$, there is not enough time for $q$ to be driven to sufficiently large values for Majorana tunnelling to take place. In the high current regime, $I_{DC}\gtrsim I_{\theta}, I_{AC} \neq 0$, if $I_{DC}-I_{AC}>I_{\theta}$ then $I(t) > I_{\theta}$ for all $t$, whilst if $I_{DC}-I_{AC}<I_{\theta}$ then, as in the intermediate regime, $I(t) \lessgtr I_{\theta}$ depending on the value of $t$. The crucial difference between this and the intermediate regime is that, since $I_{DC}\gtrsim I_{\theta}$, even as $f \rightarrow \infty$ the quasicharge is still driven to large enough values for Majorana tunnelling to take place and so, unlike the intermediate regime, there is no cut-off frequency. Note that, whilst there are quantitative differences in the behaviour of $I$ vs. $f$ for the cases $I_{DC}-I_{AC}>I_{\theta}$ and $I_{DC}-I_{AC}<I_{\theta}$, there is no qualitative distinction between them and so we do not divide the high bias current regime along these lines. To reiterate, the existence of three separate regimes is not so much a result of the value of $I(t)$ at different times, but rather the evolution of $q$ at different frequencies.

\begin{figure}
	\centering
	\includegraphics[width=\columnwidth]{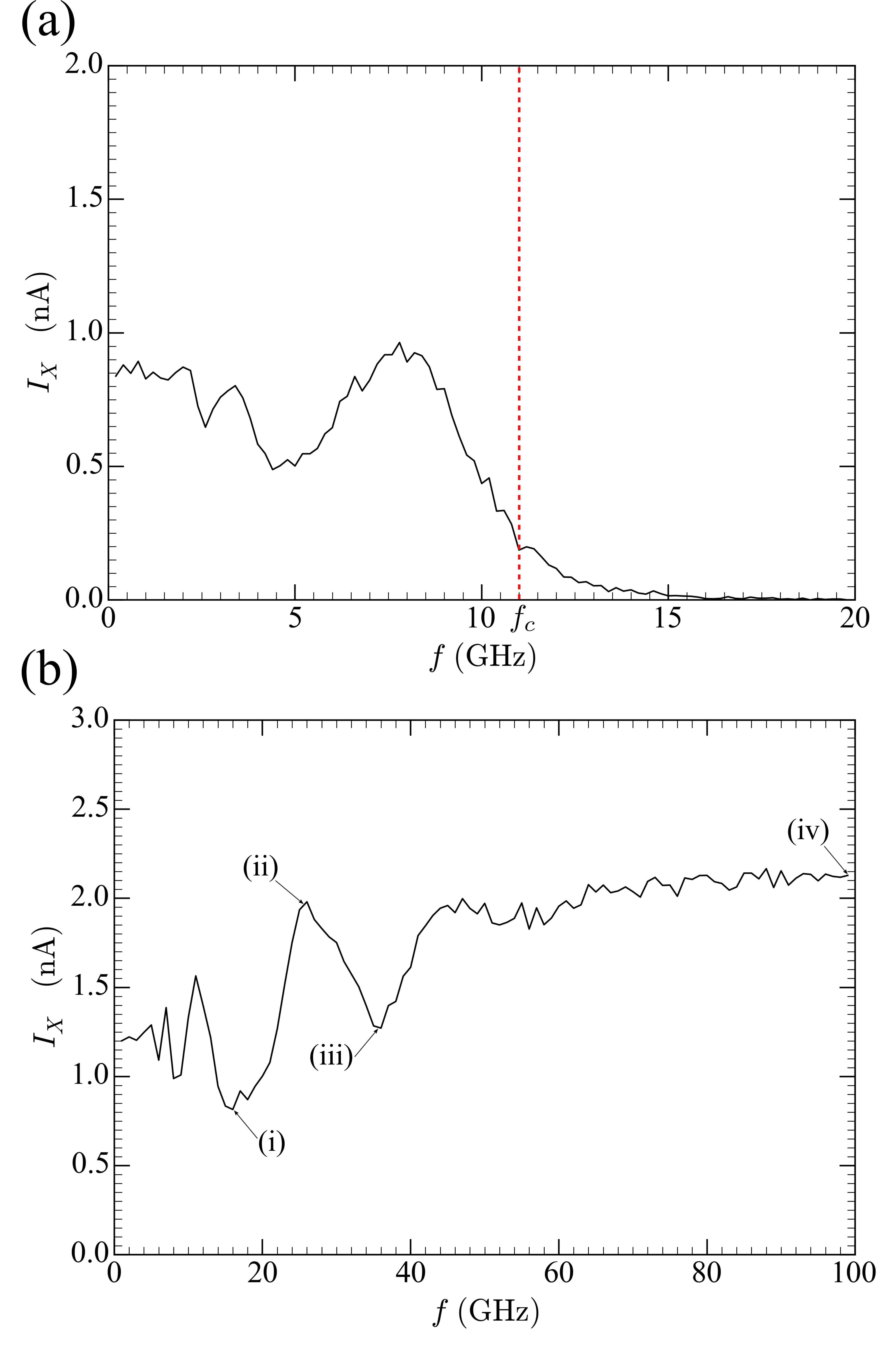}\caption{\label{IvsF} Transverse current, $I_X$,  as a function of bias current frequency for two different regimes. In both cases $I_\theta = 3.8 \text{nA}$. (a) $I_{DC} = 0.8\text{nA}\ll I_{\theta}, I_{DC} + I_{AC} = 4.8\text{nA} \gtrsim I_{\theta}$; the transverse current is finite below some threshold frequency and zero above it. (b) $I_{DC} = 4.0\text{nA} \gtrsim I_{\theta}, I_{AC} = 4.0\text{nA} \neq 0$ and $I_X$ exhibits resonances at low frequencies, before increasing to a constant value at higher frequencies. Plots of $q$ vs. $t$ at the points (i)-(iv) are shown in Fig. \ref{ringing}. In the low bias regime, $I_{DC}, I_{AC} \ll I_{\theta}$, Majorana tunnelling between the leads and TSC is negligible, resulting in $I_X \simeq 0$. For both plots a bias voltage of $V_2 = -V_1 = 0.01\text{mV}$ was used.}
\end{figure}

The behaviour of the junction voltage, $V$, and transverse current, $I_X$ in each of these three different regimes is shown in Fig. \ref{VvsF} and Fig. \ref{IvsF}. As an aside, we note that, without MBSs, there is no lead to TSC tunnelling and so $I_X =0$. The behaviour of $V$ as a function of driving frequency is similarly featureless, taking an almost constant value, $V = I_{DC}/G$, except in the case of high bias currents, such that $I_B \leq I_{DC} + I_{AC}$,  where Bloch oscillations lead to a suppression of $V$ at low frequencies. In any case, the richness of behaviour seen in Fig. \ref{VvsF} and Fig. \ref{IvsF} is absent in the topologically trivial case.

In the limit of low bias current, $I_{DC}, I_{AC} \ll I_{\theta}$, the quasicharge takes a value $q\ll e/2$, at all times and so Majorana tunnelling is negligible. This immediately implies that the transverse current will vanish, $I_X = 0$, and also permits an analytic description of the junction voltage. Solving Eq. \eqref{qevolution}, we find that, for a particular driving frequency, $f$, and at time, $t$, the junction voltage is given by,
\begin{equation}\label{approxV}
\begin{split}
&V =\frac{I_{DC}}{G}\\
& +2E_CI_{AC}e^2\left[\frac{2GE_C\cos(2\pi ft)+2\pi fe^2\sin(2\pi ft)}{\left(2GE_C\right)^2+\left(2\pi fe^2\right)^2}\right],
\end{split}
\end{equation}
where we have suppressed a rapidly decaying exponential transient term that depends on initial conditions. Time averaging this expression over a long period gives the $f$-independent result $\left<V\right> = I_{DC}/G$. The $f$ dependence of the system can instead by observed by considering the variance, $\sigma_V^2$. The solid black line in Fig. \ref{VvsF}(a) is a plot of $\sigma_V^2$ generated by simulation and is plotted along with the analytic result (dashed red line),
\begin{equation}
\sigma_V^2 \simeq \frac{2\left(E_CI_{AC}\right)^2}{\left(2GE_C\right)^2+\left(2\pi f e^2\right)^2},
\end{equation}whose derivation is detailed in Appendix \ref{voltage analytic}. As can be seen, there is very good agreement between theory and simulation, which is unsurprising since the system is in the deterministic, low bias, regime.

If the DC component of the bias current, $I_{DC}$ is much less than the threshold current, $I_{\theta}$, but the sum of the DC and AC components, $I_{AC}$, is greater than or similar to $I_{\theta}$, then the total bias current applied to the Josephson junction will oscillate between values greater and less than the threshold current. By definition of $I_{\theta}$, when $I>I_{\theta}$ the quasicharge is driven to larger values, whilst when $I<I_{\theta}$, the quasicharge tends towards its fixed value. For Majorana tunnelling to take place, it is necessary that $I>I_{\theta}$ for long enough for the quasicharge to evolve to a value $q\gtrsim e/2$. Majorana tunnelling therefore occurs at low frequencies, but ceases above some cutoff frequency, $f_c$. This is clearly shown by the behaviour of $I_X$ in Fig. \ref{IvsF}(a), where $I_X = 0$ corresponds to no Majorana tunnelling. An approximate value for $f_c$ can be calculated by considering the evolution of $q$ according to Eq. \eqref{qevolution}. As described in Appendix \ref{cut-off frequency}, we find that,
\begin{equation}\label{fcutoff}
f_c = \frac{1}{2\pi e^2}\left[\left(\frac{eI_{AC}}{\frac{q_c}{e}-\frac{eI_{DC}}{2GE_C}}\right)^2-\left(2GE_C\right)^2\right]^\frac{1}{2},
\end{equation}
where $q_c$ is the smallest magnitude of quasicharge for which Majorana tunnelling occurs at a significant rate. Taking $q_c = 0.4e$ and using the same system parameters as in Fig. \ref{IvsF}(a), the above formula predicts $f_c = 11\text{GHz}$, which we see is in reasonable agreement with the simulation. Note also that, just below $f_c$, there is a distinctive peak in $I_X$. By considering the evolution of $q$, one can understand this as corresponding to the driving frequency which is high and so rapidly brings $q$ to values near $e/2$, resulting in tunnelling, but is not so high as to cause cut-off. In plot (b) of Fig. \ref{VvsF} we see that, like the transverse current, the junction voltage adopts a constant value above some cut-off frequency. This behaviour can be understood in essentially the same terms as just described for $I_X$: at high frequencies there is no Majorana tunnelling and so, after time averaging, $\left<V\right> = I_{DC}/G$, in accordance with Eq. \eqref{approxV}; below $f_c$ Majorana tunnelling results in an average value of $q$, and therefore $V$, of close to zero.

In the large bias current limit, $I_{DC}\gtrsim I_{\theta}, I_{AC} \neq 0$, there is no frequency at which Majorana tunnelling does not take place, and therefore no cutoff frequency. However, the AC component still has an effect on $I_X$ and $V$, as shown in Fig. \ref{IvsF}(b) and Fig. \ref{VvsF}(c). Considering first the behaviour of the transverse current, we see that at high frequencies $I_X$ adopts an approximately constant value, whilst at lower frequencies it behaves highly non-monotonically. In particular, $I_X$ exhibits suppressions at the frequencies $f_s=\frac{n}{\tau}$, where $\tau$ is the average time between Majorana tunnelling events and $n$ is an integer. We note that, whilst one can, in principle, formulate an analytical expression for $\tau$, the stochastic nature of tunnelling means that, in practice, good agreement between the calculated and observed $f_s$ is found only when $\tau$ is determined by numerical simulation.

\begin{figure}
	\centering
	\includegraphics[width=214pt]{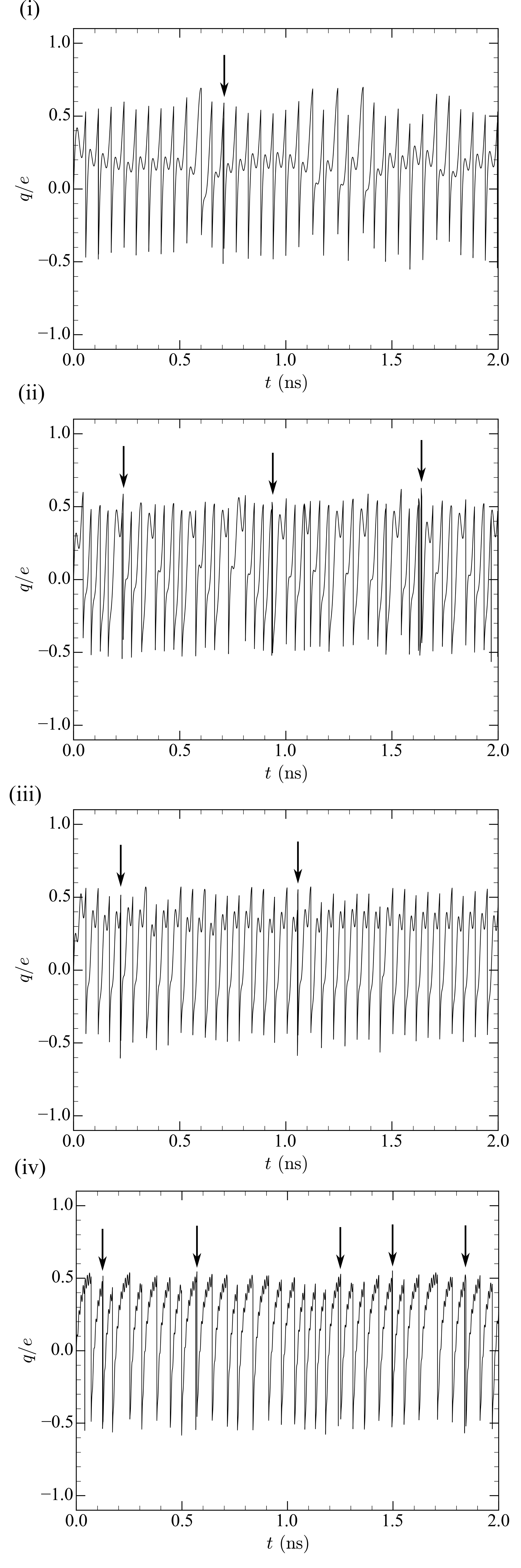}\caption{\label{ringing} Typical plots of $q$ vs. $t$ at the frequencies identified in Fig. \ref{IvsF}(b). Ringing events, indicated by arrows, are difficult to distinguish from normal tunnelling at this scale, but a clearer comparison is show in Fig. \ref{ringzoom}. Note that at the unsuppressed points, (ii) and (iv), there are more ringing events than at the suppressed points, (i) and (iii).}
\end{figure}

\begin{figure}
	\centering
	\includegraphics[width=235pt]{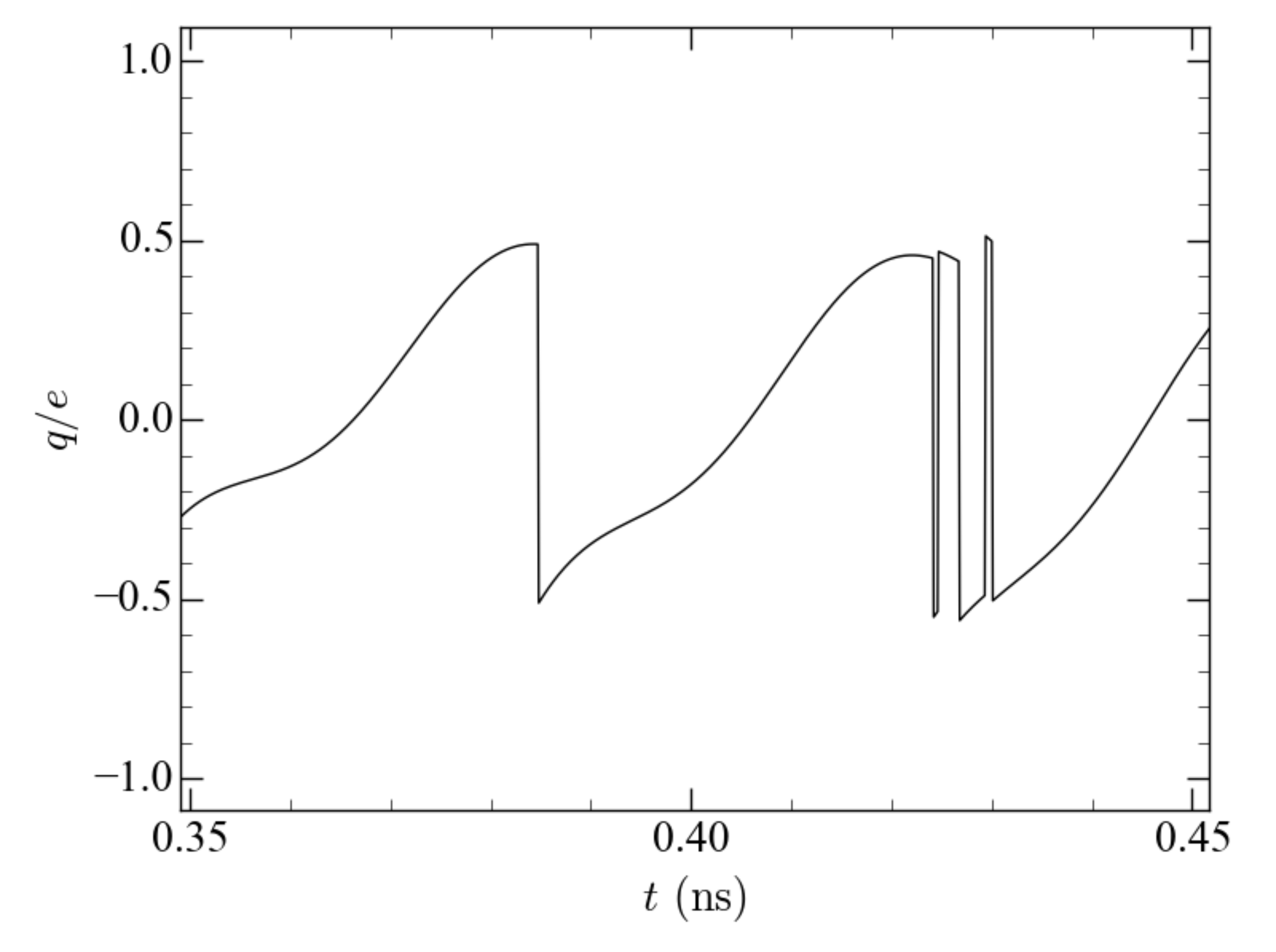}\caption{\label{ringzoom} A detailed comparison of single Majorana tunnelling and ringing. A single Majorana tunnelling event takes place at $t\approx 0.38$ns, whilst a ringing event can be seen at $t\approx 0.43$ns. The ringing event constitutes five single tunnelling events over an interval of approximately $0.01$ns, whilst the usual interval between single tunnelling events for the setup shown is $0.04$ns. Ringing therefore has a very significant impact on the total charge transferred over a given time period, and therefore the average value of $I_X$. }
\end{figure}

To understand the origin of the suppressions of $I_X$ at $f=\frac{n}{\tau}$, we must first appreciate what processes contribute to Majorana tunnelling and how these are affected by changes in the driving frequency. For the probability of a tunnelling event occurring to be non-negligible, $q$ must have a sufficiently large value (typically $|q|\gtrsim \frac{e}{2}$). This value can come about in two ways: the quasicharge is driven by $I(t)$; a Majorana tunnelling event causes $q$ to jump. In Fig. \ref{ringing} we plot $q$ vs $t$ for different driving frequencies, corresponding to the suppressions and non-suppressions seen in Fig. \ref{IvsF}(b). From the plots in Fig. \ref{ringing} it is clear that, whilst there is some variation, $f$ has relatively little impact on $\tau$, the time taken for $q$ to be driven from $-e/2$ to $+e/2$. However, one should not infer from this that $I_X$ is the same at all four frequencies since, whilst $\tau$ is relatively unchanged, there are significant differences in the number of Majorana tunnelling events that occur after $q$ has been driven into the tunnelling regime. In plots (i) and (iii), we see that Majorana tunnelling events tend to occur singly, but in plots (ii) and (iv) there is a clustering of tunnelling events such that, $I_X$ is higher in both cases, compared with (i) and (iii). This ``ringing" phenomenon where, instead of a single tunnelling event, several occur over a very short interval, is a result of jumps in $q$ repeatedly causing $|q|$ to be sufficiently large for tunnelling to take place. Although the ringing phenomenon indicated by arrows in Fig. \ref{ringing} is difficult to see, due to the very short time scale over which it takes place compared to normal tunnelling, a higher resolution comparison of ringing and single tunnelling events is shown in Fig. \ref{ringzoom}, where the single tunnelling events that make up the ringing are clearly visible. We note that Fig. \ref{ringzoom} does not take into account possible memory effects, as described in Section II.C, which may be of some importance, but for the reasons explained there we do not anticipate these effects making a significant qualitative difference to our results.

Ringing is suppressed if $I(t)$ rapidly drives the quasicharge to the region $|q|\ll \frac{e}{2}$ after a tunnelling event has taken place. Suppression of ringing therefore corresponds to $I(t)$ taking its maximum value immediately after a tunnelling event, i.e.\ we require that $\tau = \frac{n}{f_s}$, which is exactly the relation between $f_s$ and $\tau$ observed in our simulations. In addition to $\tau = \frac{n}{f_s}$, suppression of ringing also requires a specific phase relationship between $I(t)$ and the quasicharge oscillations.
However, this phase locking occurs naturally and so even if we randomize the initial phase offset for each frequency instance, as in Fig. \ref{IvsF}(b), suppression of ringing, and therefore $I_X$, is still observed.
From the simulations we see that the reason for the phase locking is that a positive or negative phase offset leads to a shorter or longer time $\tau$ to the next tunnel event, respectively. Thus each Majorana tunnelling reduces the offset and the latter vanishes after a few events. It follows that the observed $I_X$ vs. $f$ characteristics of the system are independent on the initial configurations.

It is also important to note that, even if the condition $f = \frac{n}{\tau}$ is satisfied, ringing will not be suppressed for high $f$, since each drive cycle will be too fast for $q$ to be changed significantly. Quantitatively, we expect that, for $f \gg \left(I_{DC} + I_{AC}\right)/e$, $I_X$ will be approximately constant. This effect can be seen in Fig. \ref{IvsF}(b).  Although the suppression of ringing is the main contributor to the changes in $I_X$ seen in Fig. \ref{IvsF}(b), variation of $\tau$ also has a minor effect at some frequencies. This variation in $\tau$ is a result of $I(t)$ changing the average value of $q$ over one quasicharge cycle and therefore affecting the average of $\dot{q}$ via the $G$ term in Eq. \eqref{qevolution}.  For example, comparing plots (a) and (b) in Fig. \ref{ringing}, we see that at the suppression point $f = 16.7\text{GHz}$, we obtain $\tau \approx 30 \text{ps}$, whilst in between suppressions, at $f=26.3 \text{GHz}$, we obtain $\tau \approx 25 \text{ps}$. From this change in $\tau$ alone, we would expect the suppressed value of $I_X$ to be around $80 \%$ of the unsuppressed value, but since it is actually only $40\%$, suppression of ringing is clearly a more important factor. Note also that, at $f=100 \text{GHz}$, we once again obtain $\tau \approx 30 \text{ps}$, further emphasising that changes in $\tau$ are not as important as changes in the incidence of ringing as far as suppression of $I_X$ is concerned. Panel (c) of Fig. \ref{VvsF} shows that the junction voltage, $V$, changes with $f$ in a similar manner to $I_X$. The origin of this behaviour can be seen by examining a plot of $q$ vs. $t$ at different bias frequencies, from which it is clear that, at suppression points, $q$ is driven rapidly from small values to the tunnelling regime, resulting in low average $q$, and therefore $V$. The mechanism which causes suppression of $V$ is very different to the process described above that gives rise to a suppression of $I_X$. This is because, whilst ringing makes a major contribution to the transverse current, its effect on the average value of $q$, and therefore $V$, is very similar to that of normal Majorana tunnelling, since ringing is such a rapid process. Consequently, the values of $f_s$ for $V$ are not equal to the $f_s$ for $I_X$.
\newline

As a final comment on the time-dependent driving phenomenology of the Majorana-Josephson system, it is worth noting that Fig. \ref{VvsF} and Fig. \ref{IvsF} depict changes in $V$ and $I_X$ over a frequency interval of the order of a few GHz, which may be at the limit of experimental accessibility. This is a direct consequence of the set of system parameters we have chosen to use in our simulations, in particular the values $\Gamma_{1,2}=10^{11} \text{s}^{-1}$ and $G=e^2/h$, corresponding to what we expect for typical experimental setups. If, for example, we were to instead consider a system with the less typical, but still experimentally achievable, parameter values of $\Gamma_{1,2}=10^{8} \text{s}^{-1}$ and $G=0.001e^2/h$, and decrease the magnitude of the bias current by a corresponding amount, then we would find that Fig. \ref{VvsF} and Fig. \ref{IvsF} were reproduced, but over a scale of MHz rather than GHz, and with the magnitude of $I_X$ reduced by the same factor. The phenomenology, however, does not change. We therefore see that, since $\Gamma_{1,2}$ and $G$ can be modified by careful gating of the system, the ability to measure at GHz frequencies is not required to observe the phenomena reported in this section.

% ----------------------------------------------------------------------------
\section{Conclusions}

In this work, we have demonstrated how the presence of Majorana bound states in topological superconductors can enrich the behaviour of capacitive Josephson junctions. By enabling single-particle sub-gap tunnelling between the superconductor and its surroundings, MBSs allow the Josephson junction to be perturbed in a manner not consistent with the system's underlying periodicity, and thus to be excited to a non-equilibrium state. The resulting charge dynamics of the Majorana-Josephson system are dependent upon a variety of factors, but the essential parameters are the tunnelling rate between the superconductor and metallic leads and the magnitude and time dependence of the bias current applied to the Josephson junction. We have shown that, for a static bias current, the Majorana-Josephson system may be in one of three regimes, determined by tunnelling rate and current magnitude. If the bias current is sinusoidally varying, then the system's behaviour is a function of the current frequency in a way that depends upon the current magnitude.

The charge dynamics can be observed experimentally through measurement of the voltage across the Josephson junction, as in the non-topological case, or by studying the transverse current through the Majorana-Josephson device, the existence of which is made possible by the presence of electronic states corresponding to a delocalized pair of MBSs. In either case, we have demonstrated how experimental results can be directly linked to quasicharge behaviour.

In summary, Majorana-Josephson devices represent an unusual arena in which to realize stochastic, non-equilibrium behaviour, made possible by the unique properties of Majorana bound states. Observation of the phenomena highlighted in this work would constitute a dramatic example of a macroscopic quantum effect.

% ----------------------------------------------------------------------------
\section*{Acknowledgements}

We thank R. Egger and A. Zazunov for a fruitful exchange of ideas at the early stages of this project.
B.B. and I.J.vB. thank the Departamento de F\'{i}sica Te\'{o}rica de la Materia Condensada at the Universidad Aut\'{o}noma de Madrid for hospitality. I.J.vB. acknowledges studentship funding from EPSRC under grant No. EP/I007002/1. A.L.Y. acknowledges support from the Spanish MINECO via grants FIS2014-55486P, FIS201784860-R and from the ``Mar{\'i}a de Maeztu'' Programme for Units of Excellence in R\&D (MDM-2014-0377). Open data compliance: This work is theoretical and all plots are reproducible with given formulas and parameter values.

%-----------------------------------------------------------------------------
\appendix
\section{System Parameters}\label{system parameters}

\noindent Unless explicitly noted otherwise, the following parameters were used to produce all plots shown in this work:
\newline

\begin{tabular}{ccc}
 	$T$ & $=$ & $0.05$K \\
 	$E_J$ & $=$ & $0.02$meV \\
 	$E_C$ & $=$ & $0.1$meV \\
 	$G$ & $=$ & $e^2/h$ \\
 	$\Gamma_{1,2}$ & $=$ & $10^{11} \text{s}^{-1}$ \\
 	$V_{1,2}$ & $=$ & $0$ \\
\end{tabular}
\newline

\noindent Furthermore, whenever a time average of $V$ or $I_X$ was carried out, the average was performed over an interval of $0.1\mu\text{s}$.

% ----------------------------------------------------------------------------
\section{Inter-band Transitions}\label{inter-band transitions}

\noindent There are two main mechanisms via which the Majorana-Josephson device might be excited to the second, or higher, energy bands in quasicharge space shown in Fig. \ref{bandstructure}. The first of these is straightforward thermal excitation, which has the usual probability,
\begin{equation}
P_T = \exp\left(-\frac{E_g}{k_BT}\right),
\end{equation}
where $E_g$ is the energy gap between the first and second bands. For the system parameters considered throughout this work, we find that, even at $q=\pm e$ where $E_g$ takes its lowest value $E_g \approx E_J$, the excitation probability is only $P \approx 0.01$. This indicates that thermal excitation is likely to have a negligible impact and we therefore do not consider its influence in our analysis of the Majorana-Josephson device.

\noindent Of potentially greater significance for inter-band transitions is Landau-Zener (LZ) tunnelling. The probability of this leading to an inter-band transition is \cite{geigenmuller1988},
\begin{equation}
P_Z \simeq \exp\left(-\frac{\pi e E_J^2}{\hbar E_C |I|}\right),
\end{equation}
where $|I|$ is the bias current applied across the Josephson junction. Substituting in our system parameters, we find that $P_Z \ll 1$ only for $|I| \lesssim 0.1 \text{nA}$. Hence, the rate of LZ tunnelling is appreciable in our system for most bias current values considered in this work. Despite this, we suggest that LZ tunnelling can be neglected in a description of the charge dynamics of the system. Our reasoning behind this suggestion is that, whilst LZ tunnelling does mediate an inter-band transition in the vicinity of the quasicharge zone boundary, it does not change the value of the quasicharge. The effect of LZ tunnelling is just to increase the rate of Majorana tunnelling described by Eq. \eqref{tunnelrate}, since the transition to a higher energy band results in subsequent Majorana tunnelling events having a more negative $\delta E_{ch}$. However, at low temperatures, $\exp\left(\frac{\delta E_{ch}}{k_BT}\right) \simeq 0$ near the zone boundary, even in the lowest band and so the slightly enhanced tunnelling rate due to LZ transitions to the higher band is of little relevance.

\section{Numerical Method}\label{numerical method}

As explained in Secion II, the dynamics of the Majorana-Josephson system can be understood with reference to the
quasicharge, $q(t)$, and so the numerical approach is essentially concerned with finding this quantity for a given set of system parameters and then using $q(t)$ to find any other variable desired.

The dynamics of $q(t)$ are described by Eq. \eqref{qevolution} and we integrate it using the standard Runge-Kutta 4th order algorithm. To obtain $E_0(q)$, on the right hand side of Eq. \eqref{qevolution}, we must find the ground state energy of the Hamiltonian, $H_{sc}$, given by Eq. \eqref{hsc}. As $H_{sc}$ is of the Bloch form, we take as our ansatz the wave function,

\begin{equation}
	\Psi \left(\phi, q\right) = \sum_m a^{(q)}_m e^{i\phi \left(\frac{q}{2e}+m\right)},
\end{equation}
with $m \in \mathbb{Z}$. Substituting this into the Schr\"{o}dinger equation with $H_{sc}$ gives,
\begin{equation}
\begin{split}
\sum_m \left\{-4E_C \left(\frac{q}{2e} + m\right)^2 a^{(q)}_m e^{i\phi \left(\frac{q}{2e}+m\right)} \right.\\
+ \frac{E_J}{2}\left( a^{(q)}_{m} e^{i\phi \left(\frac{q}{2e}+m+1\right)} + E_J a^{(q)}_{m} e^{i\phi \left(\frac{q}{2e}+m-1\right)} \right)\\
\left.+ Ea^{(q)}_m e^{i\phi \left(\frac{q}{2e}+m\right)}\right\} = 0.	
\end{split}
\end{equation}
Relabelling indices as appropriate and requiring that each $e^{i\phi \left(\frac{q}{2e}+m\right)}$ vanishes, we find,
\begin{equation}\label{mrequirement}
	-4E_C\left(\frac{q}{2e} + m\right)^2 a^{(q)}_m + \frac{E_J}{2}\left(a^{(q)}_{m-1}+a^{(q)}_{m+1}\right) + Ea^{(q)}_m = 0,
\end{equation}
which represents an infinite set of simultaneous equations. Note that, since the potential term in $H_{sc}$ is proportional to $\cos(\phi)$, $a^{(q)}_m$ couples only to $a^{(q)}_{m\pm 1}$.  It turns out that the truncation, $-3\leq m \leq 3$ is a very good approximation for our purposes. The energy of the lowest band, $E_0 (q)$, can then be found by computing the lowest eigenvalue of the $7\times 7$ matrix corresponding to Eq. \eqref{mrequirement} for values of $q$ in the range $-e<q\leq e$. 
 
The smooth evolution of $q(t)$ by Eq. \eqref{qevolution} is interrupted by the sudden charge jumps caused by the tunnelling into and out of the MBSs. We therefore supplement the equation of motion by checking for single particle tunnelling during each time step of the integration, which is done in the usual way by comparison of a random number in the interval $[0,1]$ with $\Gamma_{MBS} \Delta t$, where $\Delta t$ is the integration time step. Since tunnelling may occur through both leads simultaneously two independent checks are performed for the left and right leads. Furthermore Bloch reflections and oscillations are implemented when appropriate. 

Having established $q(t)$ the junction voltage is found through the relation $V = dE_0 /dq$. The transverse current
$I_X$ is calculated by a minor addition to the Runge-Kutta algorithm which counts the net flow of charge between the metallic leads through the TSC.

\section{Analytic Expressions for $q$ and $V$ with Time-varying Driving in the Low Bias Current Regime}\label{voltage analytic}

\noindent In the limit of low bias currents, $I_{DC},I_{AC}\ll I_{\theta}$, Majorana tunnelling is negligible and so the evolution of $q$ is determined entirely by Eq. \eqref{qevolution}. Furthermore, since $q\ll e$, the dispersion of the lowest energy band can be accurately modelled as $E_0 = \frac{E_C}{e^2}q^2$ and so the evolution of $q$ is described by the equation,
\begin{equation}
\dot{q} = I_{DC} + I_{AC}\cos\left(2\pi ft\right)-\frac{2GE_C}{e^2}q.
\end{equation}
The solution of this equation is elementary and gives $q(t)$. Then, using the fact that the voltage across the Josephson junction is given by $V = \frac{\text{d}E_0}{\text{d}q} \simeq \frac{2E_C}{e^2}q$, we find that,
\begin{equation}\label{appendixB}
\begin{split}
V = \frac{2E_C}{e^2}\left\{I_{AC}e^2\left[\frac{2GE_C\cos(2\pi ft)+2\pi fe^2\sin(2\pi ft)}{\left(2GE_C\right)^2+\left(2\pi fe^2\right)^2}\right]\right.\\
+\left[q_0-\frac{I_{DC}e^2}{2GE_C}-\frac{2I_{AC}e^2GE_C}{\left(2GE_C\right)^2+\left(2\pi fe^2\right)^2}\right]e^{-\frac{2GE_C}{e^2}t}\\
\left. +\frac{I_{DC}e^2}{2GE_C}\right\},
\end{split}
\end{equation}
with the exponential term being suppressed in Eq. \eqref{approxV} since it decays rapidly for typical system parameters where $\frac{GE_C}{e^2} \sim 10^{10} \text{s}^{-1}$. Over a long time interval (usually more than $100$ns) the sinusoidal and exponential terms in Eq. \eqref{appendixB} average to zero and we are left simply with the DC term,
\begin{equation}
\left<V\right>_{\delta t\rightarrow \infty} \simeq \frac{I_{DC}}{G},
\end{equation}
which is independent of frequency. However, if instead we consider the variance of the average junction voltage, $\sigma_V^2 = \left<V^2\right>-\left<V\right>^2$, then we find that
\begin{equation}
\sigma_V^2 \simeq \frac{2\left(E_CI_{AC}\right)^2}{\left(2GE_C\right)^2+\left(2\pi f e^2\right)^2},
\end{equation}
where we have neglected the rapidly decaying exponential terms in Eq. \eqref{appendixB}. We therefore see that, whilst the junction voltage itself is frequency independent, the variance in the junction voltage has a driving frequency dependence which can be measured experimentally.

\section{Cut-off Frequency in the Intermediate Bias Current Regime}\label{cut-off frequency}

\noindent Simulations demonstrate that, in the intermediate bias current regime, $I_{DC}<<I_{\theta}, I_{AC} + I_{DC} \gtrsim I_{\theta}$, there exists some cut-off frequency, $f_c$, above which Majorana tunnelling is negligible, i.e. $I_x \rightarrow 0$. To find an analytic approximation for this frequency, we begin with the expression for $q$ found by solving Eq. \eqref{qevolution},
\begin{equation}\label{qoft}
\begin{split}
&q \simeq \frac{I_{DC}e^2}{2GE_C}\\
+&I_{AC}e^2\left[\frac{2GE_C\cos(2\pi ft)+2\pi fe^2\sin(2\pi ft)}{\left(2GE_C\right)^2+\left(2\pi fe^2\right)^2}\right],
\end{split}
\end{equation}
where we have suppressed the rapidly decaying exponential term. Now, for Majorana tunnelling to be negligible, we require that $q < q_c$ for all $t$, where $q_c$ is the smallest value of the quasicharge for which tunnelling takes place at an appreciable rate. In the $T \rightarrow 0$ limit, $q_c = e/2$, but at the finite temperatures typically achieved in experiments on systems of the type we consider $q_c \approx 0.4e$. We therefore proceed by differentiating Eq. \eqref{qoft} with respect to $t$ and finding the maximum value of $q$ at any time, this is given straightforwardly by,
\begin{equation}
q_{max} = \frac{I_{AC}e^2}{\sqrt{\left(2GE_C\right)^2+\left(2\pi f e^2\right)^2}} + \frac{I_{DC}e^2}{2GE_C}.
\end{equation}
Setting $q_{max} = q_c$ and solving for $f$ gives the required expression for the cut-off frequency in the intermediate bias regime,
\begin{equation}
f_c = \frac{1}{2\pi e^2}\left[\left(\frac{eI_{AC}}{\frac{q_c}{e}-\frac{eI_{DC}}{2GE_C}}\right)^2-\left(2GE_C\right)^2\right]^\frac{1}{2},
\end{equation}
as reported in Eq. \eqref{fcutoff}.

% ----------------------------------------------------------------------------

\end{document}